\documentclass{article}
\usepackage{epsfig}
\usepackage{amssymb}
\newcommand{\bfr}{\begin{flushright}}
\newcommand{\efr}{\end{flushright}}
 
\begin{document}
\title{Motion of test particles around a charged dilatonic black hole
}
\author{Takuya Maki\\
Department of Physics, Tokyo Metropolitan University,\\
Minami-ohsawa, Hachioji-shi, Tokyo 192-03, Japan\\
and\\
Kiyoshi Shiraishi\\
Akita Junior College, Shimokitade-Sakura, Akita-shi, \\Akita 010,
Japan
}
\date{Class. Quantum Grav. {\bf 11} (1994) 227--237
}
\maketitle
\begin{abstract}
We examine motion of test particles with various masses, 
electric charges and dilatonic charges in a background metric and
fields of a charged dilatonic black hole. 
doi:10.1088/0264-9381/11/1/022
\end{abstract}

\section{Introduction}
Recently, there has been much interest in the study of charged
dilatonic black holes (BHs) \cite{1,2,3,4,5}. They have diverse
connections to supergravity \cite{1}, Kaluza-Klein \cite{2}, string
\cite{3}, and conformal field theories \cite{4}.

In order to study the role of dilaton, one can consider the action with
an arbitrary value for the dilaton coupling \cite{3}:
\begin{equation}
S=\int d^4x
\frac{\sqrt{-g}}{16\pi}[R-2(\nabla\phi)^2-e^{-2a\phi}F^2]
\label{1.1}
\end{equation}
where $R$ is the curvature scalar and $\phi$ is the dilaton.
$F_{\mu\nu}$ denotes the Maxwell field strength. The Newton constant is
normalized to unity here.

The effective field theory of string theory corresponds to the case
with $a^2=1$, while Kaluza-Klein theory (in five dimensions) requires
$a^2=3$. The usual Einstein-Maxwell system can be obtained when $a^2$
is set to be zero.

Classical and quantum properties of charged dilatonic BHs have been
investigated in various aspects by many authors \cite{3,6,7,8,9,10}.
Some critical values for the coupling $a$ have been found and discussed.

In this paper, we examine motion of test particles around the dilatonic
BH, which appears in the system governed by the `interpolating' action
(\ref{1.1}). Test particles that we consider are assumed to have
`electric' and dilatonic charges, as well as arbitrary masses. Although
the analysis will be carried out mainly for a background metric and
fields of an extreme BH (defined later), the qualitative features for
general cases are similar to those for this extremal case as far as we
treat the motion of test particles outside BHs.

The organization of this paper is as follows. In the next section, we
will present the background fields and metric for a charged dilatonic
BH located at the origin. The generic motion of a test particle in the
background is also studied in the section 2. In section 3, we study the
possibility of a static equilibrium of the system consisting of the
charged dilatonic BH and a test particle. In section 4, we treat
circular motions of test particles around the BH. The scattering
problem is considered in section 5. Finally, section 6 is devoted to
conclusion and discussion. 

\section{Motion of a test particle in a charged-dilaton-BH background}
The metric for a static, spherically-symmetric charged dilatonic BH can
be obtained by solving field equations derived from the action
(\ref{1.1}). In the line-element form, it can be written as
\begin{equation}
ds^2=g_{\mu\nu}dx^\mu
dx^\nu=-\Delta\sigma^{-2}dt^2+\sigma^2\{\Delta^{-1}dr^2+r^2(d\theta^2+\sin^2\theta
d\varphi^2)\}
\label{2.1}
\end{equation}
where 
\begin{equation}
\Delta(r)=\left(1-\frac{r_+}{r}\right)\left(1-\frac{r_-}{r}\right)
\quad\mbox{and}\quad
\sigma^2(r)=\left(1-\frac{r_-}{r}\right)^{2a^2/(1+a^2)}\,.
\label{2.2}
\end{equation}

The background configuration of the classical dilaton field and gauge
potential are given by
\begin{equation}
e^{2a\phi}=\sigma^2(r)
\quad\mbox{and}\quad
A=\frac{Q}{r}dt
\label{2.3}
\end{equation}
respectively.

In these expressions, $r_+$ and $r_-$ are constants, which are related
to the mass and charges of the BH:
\begin{equation}
2M=r_++\frac{1-a^2}{1+a^2}r_-
\quad\mbox{and}\quad
Q^2=\frac{r_+r_-}{1+a^2}
\label{2.4}
\end{equation}
where $M$ is the black hole mass and $Q$ is the electric charge
of the BH. Without loss of generality, we can consider $Q$ and $a$ to
take positive values. The horizon corresponds to $r=r_+$ for this
metric. Note that the spacetime represented by (\ref{2.1}) coincides
with the Reissner-Nordstrom (RN) spacetime when $a=0$.

On the other hand, the action for a point test particle, of which
coordinates are denoted by $x^\mu$, is
\begin{equation}
S_{tp}=-\int dt\left[m
e^{b\phi}\sqrt{-g_{\mu\nu}\frac{\partial x^\mu}{\partial t}
\frac{\partial x^\nu}{\partial t}}+e A_\mu\frac{\partial
x^\mu}{\partial t}\right]
\label{2.5}
\end{equation}
where $m$, $e$ and $b$ stand for the mass, electric charge, and
coupling to the dilaton field, of the test particle, respectively.
$g_{\mu\nu}(x)$, $A_\mu(x)$, and $\phi(x)$ must be regarded as the
classical background field (\ref{2.1}), (\ref{2.3}) in the present
analysis. The radiations of electromagnetic, gravitational, and
dilatonic (scalar) waves from the particle and the associated back
reactions are therefore ignored in the present analysis.

Because of the spherical symmetry of the background fields and
spacetime, the motion of the test particle is restricted in a plane
which contains the centre of the charged dilatonic BH. We can take a
plane defined by $\theta=\pi/2$ as such a plane. There are two
constants of the motion corresponding to the Killing vectors of the
spacetime. They are
\begin{eqnarray}
& &-m e^{b\phi}g_{tt}\frac{dt}{d\tau}+e
A_t=m\Delta(r)\sigma^{(b/a)-2}(r)\frac{dt}{d\tau}+\frac{e Q}{r}\equiv E
\label{2.6}
\\
& &-m
e^{b\phi}g_{\varphi\varphi}\frac{d\varphi}{d\tau}=-m\sigma^{2+(b/a)}(r)r^2
\frac{d\varphi}{d\tau}\equiv -L
\label{2.7}
\end{eqnarray}
where 
\begin{equation}
d\tau\equiv\sqrt{-g_{\mu\nu}\frac{\partial x^\mu}{\partial t}
\frac{\partial x^\nu}{\partial t}}dt\,.
\end{equation}

The `on mass shell' condition trivially derived from the action
(\ref{2.5}), i.e.,
\begin{equation}
g^{\mu\nu}(P_\mu+e A_\mu)(P_\nu+e A_\nu)+m^2 e^{2b\phi}=0
\label{2.8}
\end{equation}
where 
\begin{equation}
P_\mu\equiv\frac{\delta S_{tp}}{\delta (dx^\mu/dt)}
\label{2.9}
\end{equation}
can be rewritten by using the background fields and the two constants as
\begin{equation}
\left(\frac{dr}{d\tau}\right)^2=\frac{\sigma^{-2b/a}}{m^2}
\left(E-\frac{e Q}{r}\right)^2-\Delta\sigma^{-2}\left(1+
\frac{L^2\sigma^{-2-(2b/a)}}{m^2r^2}\right)\,.
\label{2.10}
\end{equation}

We analyse the test-particle motion by using the equations (\ref{2.6}), 
(\ref{2.7}) and  (\ref{2.10}). The analyses for specific motions will
be treated in the subsequent sections.

\section{Static equilibrium}
First we consider the particle motion with a constant distance $r$ from
the centre of the dilatonic BH. Setting $dr/d\tau=0$ in equation
(\ref{2.10}), we obtain
\begin{equation}
\frac{E}{m}=\frac{e Q}{m r}+\sqrt{\Delta\sigma^{-2+(2b/a)}
\left(1+\frac{L^2\sigma^{-2-(2b/a)}}{m^2r^2}\right)}\,.
\label{3.1}
\end{equation}

The right-hand side of equation (\ref{3.1}) can be considered as a
function of $r$. In a stationary system, $E$ must be an extremal value.
Thus the value for $r$ that is realized in such a system is given by
the solution of the equation
\begin{equation}
\frac{dV}{dr}=0
\end{equation}
where
\begin{equation}
V(r)\equiv\frac{e Q}{m r}+\sqrt{\Delta\sigma^{-2+(2b/a)}
\left(1+\frac{L^2\sigma^{-2-(2b/a)}}{m^2r^2}\right)}\,.
\label{3.2}
\end{equation}

In this and the next sections, we analyse mainly the cases with an
extreme BH ($r_-=r_+$). The qualitative features for general cases with
non-zero charge are similar to the extremal case. In the extremal case,
the effective potential $V$ is simplified as
\begin{eqnarray}
V(r)&=&\frac{e r_+}{\sqrt{1+a^2}m
r}\nonumber \\
&+&\left(1-\frac{r_+}{r}\right)^{(1+ab)/(1+a^2)}\sqrt{1+
\frac{L^2}{m^2r^2}
\left(1-\frac{r_+}{r}\right)^{-2a(a+b)/(1+a^2)}}\,.
\label{3.3}
\end{eqnarray}

When $L\ne 0$, $r$ represents the radius of a circular orbit. This case
will be treated in the next section. In the present section, we examine
the possibility of static equilibrium by using the potential
(\ref{3.3}) with $L$ set to be $0$.

We can solve (\ref{3.1}) and (\ref{3.2}) with respect to $m$,which
takes a positive value, to obtain an allowed region in the parameter
space. For $L=0$ and an arbitrary charge of the BH, we obtain the
following inequality:
\begin{equation}
\frac{r_+(1-r_-/r_{se})}{r_-(1-r_+/r_{se})}\gtrless
\frac{a^2-2ab-1}{1+a^2}\quad\mbox{for}\quad
e\gtrless 0
\end{equation}
where $r_{se}$ indicates the distance of the equilibrium point from the
centre of the BH. For the extremal case, the inequality is simply
reduced to $1+ab>0$ for $e>0$ (and $1+ab<0$ for $e<0$).

From now on, we focus our attention to the case with the extreme BH.
For the case with a usual, extreme RN BH, which corresponds to the
case with $a=0$ here, the stable balance between gravity and Coulomb
force can never be attained except for the case with $e/m=1$. The
condition $e/m=1$ is just the extremity condition for the test
particle. In this case, the static forces are cancelled with each other
at an arbitrary distance from the extreme RN BH. In the presence of the
dilatonic force, there is a possibility that the three forces on the
test particle keep the balance at a certain distance from the hole.

Taking the stability against a small shift of the distance into
consideration, the equilibrium turns out to be realized in the three
cases for the dilaton couplings and charges of test particles as
follows:

(I) $b=a~and~e/m=(1+a^2)^{1/2}$

The test particle has like (electric) charges and its charge satisfies
the extremal condition in this case \cite{8}. The static configuration
is permitted for an arbitrary distance between the extreme black hole
and the test particle.

(II) $b>a~and~0<e/m<(1+ab)/(1+a^2)^{1/2}$

In this case, the dilatonic force is attractive while the Coulomb force
is repulsive. The distance between the black hole and the test particle
is given by
\begin{equation}
r_{se}=r_+\left\{1-\left(\frac{\sqrt{1+a^2}e}{(1+ab)m}\right)^{(1+a^2)/(a(b-a))}\right\}^{-1}\,.
\end{equation}

(III) $b<-1/a~and~e<0~and~|e|/m>|1+ab|/(1+a^2)^{1/2}$

In this case, the dilatonic force is repulsive while the Coulomb force
is attractive. The distance between the black hole and the test particle
is given by
\begin{equation}
r_{se}=r_+\left\{1-\left(\frac{\sqrt{1+a^2}|e|}{|1+ab|m}\right)^{(1+a^2)/(a(b-a))}\right\}^{-1}\,.
\end{equation}

In the cases (II) and (III), the equilibrium distances are always
outside the BH horizon ($r_{se}>r_+$), though the balances are unstable.

\section{Circular motions}
Let us consider the motion along a circular orbit with the radius $r$
around a maximally-charged dilatonic BH ($r_-=r_+$). Since it is not
pedagogical to survey all the cases in the vast range of charges of the
test particle, we will choose some particular cases for the couplings
and charges.

(I) $e=0$.

We first examine the case with $e=0$ (an electrically neutral test
particle).

The effective potential becomes
\begin{equation}
V(r)=\left(1-\frac{r_+}{r}\right)^{(1+ab)/(1+a^2)}\sqrt{1+
\frac{L^2}{m^2r^2}
\left(1-\frac{r_+}{r}\right)^{-2a(a+b)/(1+a^2)}}\,.
\label{4.1}
\end{equation}
We restrict ourselves further to two cases for the extreme BH:

(i) $a=0$. This leads to the case with an extreme RN BH. The motion of
the test particle is obviously independent of the value for its
dilatonic coupling. The effective potential has an extremum outside the
BH if
\begin{equation}
\frac{L^2}{m^2}>8r_+^2\qquad (e=0, a=0)\,.
\end{equation}
The minimum radius of the circular orbit is given when $L^2/m^2=8r_+^2$
by
\begin{equation}
r_{min}=4r_+\qquad (e=0, a=0)
\end{equation}
and then the energy of the test particle is given by
\begin{equation}
\frac{E_{min}^2}{m^2}=\frac{27}{32}\qquad (e=0, a=0)\,.
\end{equation}
In this case, the binding energy $(m-E)$ becomes about $8.1\%$ of the
rest energy.

(ii) $a=1$. This dilaton coupling corresponds to the field theory limit
of string theory. The critical behaviour is classified by the value for
$b$.

\textbullet ~$b>0$. The circular motion is possible as long as $L^2\ne
0$. The minimum radius of the circular orbit is given by $r_{min}=r_+$
in the limit of $L\rightarrow 0$. In this limit, the energy of the
particle $E$ approaches zero, i.e., the binding energy becomes $100\%$
of the rest energy!

\textbullet ~$b=0$. The circular motion is possible if
\begin{equation}
\frac{L^2}{m^2}>\frac{r_+^2}{2}\qquad (e=0, a=1, b=0)\,.
\end{equation}
The minimum radius of the circular orbit is given when $L^2/m^2=r_+^2/2$
by
\begin{equation}
r_{min}=r_+\qquad (e=0, a=1, b=0)
\end{equation}
and then the energy of the test particle is given by
\begin{equation}
\frac{E_{min}^2}{m^2}=\frac{1}{2}\qquad (e=0, a=1, b=0)\,.
\end{equation}
In this case, the binding energy $(m-E)$ becomes about $29.3\%$ of the
rest energy.

\textbullet ~$-1<b<0$. The circular motion is possible if
\begin{equation}
\frac{L^2}{m^2}>\frac{(1-b^2)r_+^2}{2}\left(\frac{-b}{1-b}\right)^{b}\qquad
(e=0, a=1, -1<b<0)\,.
\end{equation}
The minimum radius of the circular orbit is given when $L^2/m^2=r_+^2/2$
by
\begin{equation}
r_{min}=(1-b)r_+\qquad (e=0, a=1, -1<b<0)
\end{equation}
and then the energy of the test particle is given by
\begin{equation}
\frac{E_{min}^2}{m^2}=\frac{1}{2}\left(\frac{-b}{1-b}\right)^{b}\qquad
(e=0, a=1, -1<b<0)\,.
\end{equation}
In the limit of $b\rightarrow -1$, the binding energy $(m-E)$
approaches zero.

\textbullet ~$b\le -1$. The circular motion is impossible in this case
($e=0, a=1, b\le -1$).


(II) $e/m=-(1+a^2)^{1/2}~and~a=b$

In this case the effective potential becomes
\begin{equation}
V(r)=-\frac{r_+}{r}+\left(1-\frac{r_+}{r}\right)\sqrt{1+
\frac{L^2}{m^2r^2}
\left(1-\frac{r_+}{r}\right)^{-4a^2/(1+a^2)}}\,.
\label{4.2}
\end{equation}
According to the magnitude of $a$, possible circular motions are
classified:

(i) $a^2\ge 1$. The circular motion is possible as long as $L^2\ne 0$.
The radius of the circular orbit approaches $r_{min}=r_+$ in the limit
of $L\rightarrow 0$. In this limit, the energy of the test particle
$E$ approaches $-m$, i.e., the binding energy becomes $200\%$
of the rest energy!

(i) $a^2< 1$. The circular motion is possible if
\begin{eqnarray}
& &\frac{L^2}{m^2}>\frac{L_c^2}{m^2}\equiv
x^2\frac{2x-(3-a^2)/(1+a^2)}{[x-2/(1+a^2)]^2}
\left(1-\frac{1}{x}\right)^{4a^2/(1+a^2)}r_+^2\nonumber \\
& &(e/m=-(1+a^2)^{1/2}, b=a<1)
\end{eqnarray}
where
\begin{equation}
x(a)=\frac{3-a^2+\sqrt{(3-a^2)(1-a^2)}}{1+a^2}\,.
\end{equation}
The minimum radius of the circular orbit is given when $L^2=L_c^2$
by
\begin{equation}
r_{min}=x(a)r_+\qquad (e/m=-(1+a^2)^{1/2}, b=a<1)
\end{equation}
and then the energy of the test particle is given by
\begin{eqnarray}
& &\frac{E_{min}}{m}=
\frac{[(1-a^2)/(1+a^2)][3x-(3-a^2)]/(1+a^2)}{x[x-2/(1+a^2)]}\nonumber \\
& &(e/m=-(1+a^2)^{1/2}, b=a<1)\,.
\end{eqnarray}
The binding energy in this limiting case becomes about 
$13.4\%$ for $a=0$, while it becomes about $22.2\%$ for $a^2=1/3$.
In the limit of $a\rightarrow 1$, $E_{min}$ vanishes.

\section{Scattering of particles by the charged dilatonic BH}
Using (\ref{2.7}) and (\ref{2.10}), one can find the differential
equation which determines the shape of the trajectory:
\begin{equation}
\left(\frac{dr}{d\varphi}\right)^2=r^4\left[
\frac{\sigma^4}{L^2}\left(E-\frac{eQ}{r}\right)^2-\Delta
\left(\frac{m^2\sigma^{2+(2b/a)}}{L^2}+\frac{1}{r^2}\right)\right]\,.
\label{5.1}
\end{equation}

For the scattering problem, we can substitute the constants $E$ and $L$
by the value at spatial infinity as
\begin{equation}
E=\frac{m}{\sqrt{1-v^2}}\quad\mbox{and}\quad L=\frac{mvd}{\sqrt{1-v^2}}
\label{5.2}
\end{equation}
where $v$ is the velocity of the test particle at spatial infinity and
$d$ is the impact parameter. Using this relation, equation (\ref{5.1})
can be rewritten as
\begin{equation}
\left(\frac{dr}{d\varphi}\right)^2=r^4\left[
\frac{\sigma^4}{v^2d^2}\left(1-\frac{\sqrt{1-v^2}eQ}{mr}\right)^2-\Delta
\frac{(1-v^2)\sigma^{2+(2b/a)}}{v^2d^2}-\frac{\Delta}{r^2}\right]\,.
\label{5.3}
\end{equation}

Here we examine the scattering problem in a few simple situations.

(I) \textit{`Small angle scattering' of a massive test particle}
($m\ne 0$)

If the value of impact parameter is much larger than the radius of the
black hole horizon and the absolute value of the potential energy is
much smaller than the kinetic energy of the test particle, the
scattering angle is expected to be small. In such a case, the analysis
of scattering can be simplified to some extent. Equation (\ref{5.3})
can be approximated by
\begin{equation}
\left(\frac{dr}{d\varphi}\right)^2=r^4\left[
\frac{1}{d^2}\left(1+\frac{B}{r}\right)-\frac{1}{r^2}\right]
\label{5.4}
\end{equation}
where
\begin{equation}
B\equiv\frac{1}{v^2}\left\{
(1-v^2)\left(r_++\frac{1+3a^2+2ab}{1+a^2}r_-\right)-
\left(\frac{2\sqrt{1-v^2}eQ}{m}+\frac{4a^2}{1+a^2}r_-\right)\right\}
\label{5.5}
\end{equation}
is a constant, whose dimension is $[L^1]$. $|B|/d$ should be a small,
dimensionless value in the present assumption.

Equation (\ref{5.4}) can easily be solved to obtain the scattering
angle. At first order in $B$, the scattering angle is found to be
\begin{equation}
\Theta=|B~/d+O((B/d)^3)\,.
\label{5.6}
\end{equation}
 Then the differential cross-section for the small-angle scattering is
obtained as
\begin{equation}
\frac{d\sigma}{d\Omega}=\frac{d^4}{B^2}=\frac{B^2}{\Theta^4}\,.
\label{5.7}
\end{equation}
This is the same form as the Rutherford scattering at small angle.

Further, let us consider the low-velocity limit. In the case of $v^2\ll
1$, $B$ can be simplified as
\begin{equation}
B=\frac{1}{\frac{1}{2}mv^2}\
\left(Mm+\frac{ab}{1+a^2}r_--eQ\right)\quad (v^2\ll 1)
\label{5.8}
\end{equation}
where $M$ is the mass of the charged dilatonic BH. Thus we find that in
the low-velocity, or, low-energy limit the small angle scattering is
regarded as the Rutherford scattering by three inverse-square-law
forces, i.e., gravity, Coulomb force and dilatonic force.

(II) \textit{Low-energy scattering of an extreme BH and an extreme test
particle with the same dilatonic coupling} ($r_+=r_-~and~a=b~ and
e/m=(1+a^2)^{1/2}$)

As seen from the result of (I), the leading scattering amplitude could
take a velocity-independent value if the static forces are cancelled by
each other. As such a simple case, we suppose that the extreme
condition is satisfied not only among the parameters of the BH but also
among those of the test particle. Further we assume $a=b$. We restrict
ourselves to the low-velocity scattering for simplicity.

In this case, equation (\ref{5.3}) can be written as
\begin{equation}
\left(\frac{dr}{d\varphi}\right)^2=r^4\left[
\frac{1}{d^2}\left(1-\frac{r_+}{r}\right)^{(1+5a^2)/(1+a^2)}-
\frac{1}{r^2}\left(1-\frac{r_+}{r}\right)^2\right]\quad (v^2\ll 1)\,.
\label{5.9}
\end{equation}

For $a^2>1/3$, the right-hand side of equation (\ref{5.9}) becomes zero
for a certain value for $r$, which is larger than $r_+$. The value
gives the minimum distance between the centre of the BH and the
trajectory of the particle. Thus the particle can never be absorbed by
the BH in this case for $a^2>1/3$. For $a^2<1/3$, the particle can be
absorbed if the impact parameter is sufficiently small.

In general cases, one cannot express the scattering angle using
analytic functions. We show the expression for the scattering angle for
some cases with special value for $a$.

(i) $a=0$. In this case, the scattering angle is written by using the
elliptic integral:
\begin{equation}
\Theta=\frac{4}{\sqrt{(1-\beta)(\alpha-\gamma)}}\,
F\left(\sin^{-1}\sqrt{\frac{\beta(\alpha-\gamma)}{\alpha(\beta-\gamma)}}
,\sqrt{\frac{(1-\alpha)(\beta-\gamma)}{(1-\beta)(\alpha-\gamma)}}\right)-\pi\quad
(a^2=0)
\label{5.10}
\end{equation}
where $F(\varphi,k)$ is the elliptic integral of the first kind, that
is:
\begin{equation}
F(\varphi,k)\equiv\int_0^\varphi\frac{d\theta}{\sqrt{1-k^2\sin^2\theta}}
\,.
\label{5.11}
\end{equation}
$\alpha$, $\beta$, and $\gamma$ ($\alpha>\beta>\gamma$) are roots of
the following algebraic equation of third degree:
\begin{equation}
u^3-u^2+\left(\frac{r_+}{d}\right)^2=0\,.
\label{5.12}
\end{equation}
If the equation does not have real roots in the range $[0,1]$, the
trajectory of the particle ends at the horizon; the particle is
swallowed by the BH. This occurs when $d<(3\sqrt{3}/2)r_+$.

For small angle scattering, the scattering angle is approximately given
by:
\begin{equation}
\Theta=\frac{r_+}{d}\qquad\mbox{if } d\gg r_+\quad(a^2=0)\,.
\label{5.13}
\end{equation}

(ii) $a^2=1/3$. In this case, the scattering angle is written by:
\begin{equation}
\Theta=\frac{4}{\sqrt{1-\left(\frac{r_+}{d}\right)^2}}\sin^{-1}
\sqrt{\frac{1}{2}\left(1+\frac{r_+}{d}\right)}-\pi\quad
(a^2=1/3)\,.
\label{5.14}
\end{equation}
Note that $\Theta$ diverges in the limit of $d\rightarrow r_+$.

For small angle scattering, the scattering angle is approximately
by:
\begin{equation}
\Theta=\frac{2r_+}{d}\qquad\mbox{if } d\gg r_+\quad(a^2=1/3)\,.
\label{5.15}
\end{equation}

(iii) $a^2=1$. In this case, the scattering angle is simply written by:
\begin{equation}
\Theta=2\tan^{-1}\frac{r_+}{2d}\quad(a^2=1)\,.
\label{5.16}
\end{equation}
The relation is very similar to the one for the Rutherford scattering.
The differential cross section is calculated using (\ref{5.16}):
\begin{equation}
\frac{d\sigma}{d\Omega}=\frac{r_+^2}{16\sin^4(\Theta/2)}\quad(a^2=1)\,.
\label{5.17}
\end{equation}

In the present case, the small amount of the scattering angle is
independent of the small value of the incident velocity, as expected.

The scattering of two extreme BHs has been analysed by using the moduli
space metric \cite{8}. It is known that the small-mass limit for one BH
leads to the same description of scattering by the test-particle
analysis in this case (II).

(III) \textit{Massless particle} ($m=0$)

For the massless case, we must replace (\ref{5.3}) by
\begin{equation}
\left(\frac{dr}{d\varphi}\right)^2=r^4\left[
\frac{\sigma^4}{d^2}-\frac{\Delta}{r^2}\right]\,.
\label{5.18}
\end{equation}
(This is independent of the charges of test particles ($e$ and $b$).)

We again consider the extreme BH for simplicity.

In this case, equation (\ref{5.18}) can be written as
\begin{equation}
\left(\frac{dr}{d\varphi}\right)^2=r^4\left[
\frac{1}{d^2}\left(1-\frac{r_+}{r}\right)^{4a^2/(1+a^2)}-
\frac{1}{r^2}\left(1-\frac{r_+}{r}\right)^{2}\right]\quad (m=0)\,.
\label{5.19}
\end{equation}

For $a^2>1$, the right-hand side of equation (\ref{5.19}) becomes zero
for a certain value for $r$, which is larger than $r_+$. Thus the
massless particle never be absorbed by the extreme BH in this case for
$a^2>1$. For $a^2<1$, the massless particle can be absorbed if the
impact parameter is sufficiently small.

We show the expression for the scattering angle for some cases with
special value for $a$.

(i) $a^2=0$. In this case, the scattering angle is written by using the
elliptic integral:
\begin{equation}
\Theta=\frac{4}{\sqrt{(\alpha-\gamma)(\beta-\delta)}}\,
F\left(\sin^{-1}\sqrt{\frac{\gamma(\beta-\delta)}{\beta(\gamma-\delta)}}
,\sqrt{\frac{(\alpha-\beta)(\gamma-\delta)}{(\alpha-\gamma)(\beta-\delta)}}\right)-\pi\quad
(a^2=0)
\label{5.20}
\end{equation}
where $\alpha$, $\beta$, $\gamma$ and $\delta$
($\alpha>\beta>\gamma>\delta$) are roots of the following algebraic
equation of fourth degree:
\begin{equation}
u^2(u-1)^2-\left(\frac{r_+}{d}\right)^2=0\,.
\label{5.21}
\end{equation}
If the equation (\ref{5.21}) does not have real roots in the range
$[0,1]$, the trajectory of the particle ends up at the horizon; the
massless particle is swallowed by the BH. This occurs when
$d<4r_+$.

For small angle scattering, the scattering angle is approximately given
by:
\begin{equation}
\Theta=\frac{2r_+}{d}\qquad\mbox{if } d\gg r_+\quad(a^2=0)\,.
\label{5.22}
\end{equation}

(ii) $a^2=1/3$. In this case, the scattering angle is simply written by:
\begin{equation}
\Theta=\frac{4}{\sqrt{(1-\beta)(\alpha-\gamma)}}\,
F\left(\sin^{-1}\sqrt{\frac{\beta(\alpha-\gamma)}{\alpha(\beta-\gamma)}}
,\sqrt{\frac{(1-\alpha)(\beta-\gamma)}{(1-\beta)(\alpha-\gamma)}}\right)-\pi\quad
(a^2=1/3)
\label{5.23}
\end{equation}
where $\alpha$, $\beta$, and $\gamma$
($\alpha>\beta>\gamma$) are roots of the following algebraic
equation of third degree:
\begin{equation}
u^3-u^2+\left(\frac{r_+}{d}\right)^2=0\,.
\label{5.24}
\end{equation}
If the equation (\ref{5.24}) does not have real roots in the range
$[0,1]$, the
massless particle is swallowed by the BH. This occurs when
$d<(3\sqrt{3}/2)r_+$.

The trajectory takes the same form as that with $a=0$ in case (II),
examined previously.

(iii) $a^2=1$. In this case, the scattering angle is simply written
by:
\begin{equation}
\Theta=\frac{4}{\sqrt{1-\left(\frac{r_+}{d}\right)^2}}\sin^{-1}
\sqrt{\frac{1}{2}\left(1+\frac{r_+}{d}\right)}-\pi\quad
(m=0,~a^2=1)\,.
\label{5.25}
\end{equation}
The trajectory takes the same form as that with $a=0$ in the case
with $a^2=1/3$ in case (II), examined previously.

(iv) $a^2=3$. Then the massless particle is scattered in the manner of
the Rutherford scattering:
\begin{equation}
\Theta=2\tan^{-1}\frac{r_+}{2d}\quad(m=0,~a^2=3)\,.
\label{5.26}
\end{equation}
The differential cross section is:
\begin{equation}
\frac{d\sigma}{d\Omega}=\frac{r_+^2}{16\sin^4(\Theta/2)}
\quad(m=0,~a^2=3)\,.
\label{5.27}
\end{equation}

\section{Conclusions and discussions}
In this paper, we have examined the motion of test particles in the
background metric and fields of a charged dilatonic BH. As shown in
section 3, there is the static equilibrium between a test particle and
an extreme dilatonic BH in an arbitrary distance under the condition
that $a=b$ and $e/m=(1+a^2)^{1/2}$. This is similar to the RN BH case.
We also found that the static equilibrium is realized at a certain
distance determined by $e/m$ and the dilaton coupling $a$ for the
extreme dilatonic BH. Furthermore we examined the circular motion
around the extreme dilatonic BH and found that the marginally stable
radius for $a=1$ becomes smaller than for corresponding RN BH with the
same mass $M$ and that the binding energy can become large, e.g.,
$100\%$ for $a=1$ and $b>0$, $200\%$ for $e/m=(1+a^2)^{1/2}$ and
$a=b>1$. For the cases one can expect more energy release from the
dilatonic BH than that from the RN BH if there is the relevant
mechanism like the Penrose process or superradiance. In section 5 we
studied the scattering of a test particle by an extreme BH. The amount
of the lensing effect depends on the dilatonic coupling $a$. The amount
of the lensing effect becomes larger for $a=1$, smaller for $a^2=1/3$
than that for the RN BH. These facts on the dilatonic force would be of
interest from cosmological and astrophysical view points, if the
massless dilaton has been existed in the very early universe.

Although we have treated some restricted cases, the feature of the
test-particle motion in general situations could be investigated in
similar methods. One can analyse the test particle motion around the
extended object of the other types \cite{5}, similarly in the
pedagogical manner we have examined in the present paper. Besides, it
will provide a simple and effective method to reveal critical values
for parameters in such models.


\end{document}